# Bridging Design and Execution: A Visual Graph Editor for Edge and Cloud Workflows

Katarina-Glorija Grujić[1[0000-0003-2816-7980]], Nikola Stanković[2[0009-0001-8484-5478]], Maja Vukasović[2[0000-0003-0647-1922]] and Miloš Simić[1[0000-0001-8646-1569]]

[1] Faculty of Technical Sciences, University of Novi Sad
[2] School of Electrical Engineering, University of Belgrade

**Abstract.** Designing modular applications for edge and cloud computing environments involves coordinating multiple computational kernels, shared data, and event-driven execution. This paper presents a domain-specific visual graph editor that enables users to model such applications using a unified interface. The editor supports three first-class abstractions: kernels, representing computational units; shared memory nodes, modeling distributed data; and event triggers, capturing execution dependencies. Users can construct graphs visually, configure node properties, and connect elements to define data and control flow. The resulting graphs are automatically serialized into machine-readable representations (JSON/XML) and can be passed to an execution API, bridging the gap between design-time modeling and runtime deployment. The editor's graph-based model improves reasoning about data sharing, execution order, and dependencies, particularly in distributed edge and cloud scenarios. To demonstrate its applicability, we discuss a federated learning workflow, where local training kernels interact with a shared global model through event-driven coordination. Compared to traditional workflow editors and general-purpose diagramming tools, the proposed system provides explicit execution semantics, modularity, and direct deployability. This work lays the foundation for visual orchestration of modular computation in distributed environments and offers extensibility for user-defined kernels, event types, and alternative execution backends, enabling future exploration of complex distributed applications.

## 1 Introduction

Edge and cloud computing environments increasingly rely on modular, distributed computation, where applications are composed of multiple interacting components rather than monolithic services [1, 2, 3, 4]. These components—often referred to as kernels—communicate through shared data, trigger each other's execution via events, and are deployed across heterogeneous infrastructure. While such modularity improves flexibility and scalability, it also introduces significant complexity in orchestration, coordination, and reasoning about data sharing and execution flow [5, 6].

Existing orchestration approaches typically rely on low-level configuration files, workflow descriptions, or infrastructure-centric abstractions [5, 7]. Although general-purpose diagramming tools (e.g., draw.io) and workflow editors can visually represent dependencies, they lack domain-specific semantics for kernel communication, shared distributed memory, and event-driven execution [8, 9, 10]. As a result, the visual models created with such tools remain disconnected from execution, requiring additional manual translation into deployable specifications.

This paper presents a specialized graph-based visual editor designed for modeling kernel-based applications in edge and cloud computing environments. The proposed editor enables users to visually construct executable application graphs by combining three first-class entities: kernels, shared memory nodes, and event triggers. Kernels represent computational units defined externally by users; memory nodes explicitly model shared distributed data; and event triggers capture execution dependencies within the system [11]. Directed edges define communication, data flow, and control relationships between these entities.

Unlike general-purpose graph editors, the proposed system associates each graph element with configurable properties, such as kernel parameters or memory identifiers. The resulting graph is not merely illustrative: it is automatically parsed into a machine-readable representation (e.g., JSON or XML) that can be passed to an execution API. This representation preserves both structural and semantic information, enabling the runtime system to instantiate kernels, manage shared distributed memory, and coordinate execution according to the modeled graph [11].

By making memory and events explicit graph elements, the editor supports clearer reasoning about data sharing, execution order, and dependencies, which is particularly important in edge and cloud scenarios where resources are distributed and loosely coupled [1, 2]. The visual approach lowers the barrier for designing complex modular applications while maintaining a direct path from design to execution.

The contributions of this paper are:

1. A domain-specific visual graph editor for modeling kernel-based applications in edge and cloud computing environments.
2. A unified graph model that explicitly represents computation (kernels), shared distributed memory, and event-driven execution.
3. A foundation for future extensibility, allowing user-defined kernels, configurable events, and alternative execution backends.

The remainder of this paper is organized as follows: Section 2 discusses related work in visual orchestration and workflow modeling. Section 3 introduces the graph model and its semantics. Section 4 describes the user interface and interaction design of the editor. Section 5 discusses limitations and future directions, and Section 6 concludes the paper.

## 2 Related work

Several prior works address visual modeling, workflow orchestration, and distributed execution, but few provide domain-specific abstractions for kernel-based computation with explicit shared memory and event-driven semantics.

Traditional diagramming tools such as draw.io or Microsoft Visio allow users to construct general-purpose graphs but lack execution semantics [6]. Workflow management systems and orchestration frameworks, including Apache Airflow and Argo Workflows, provide mechanisms for scheduling and managing data dependencies, yet they do not expose shared memory as a first-class construct or provide a visual authoring interface tightly coupled to execution [5-7].

Edge and microcloud orchestration platforms, such as K3s and lightweight container-based systems, support deployment of modular applications on distributed nodes but typically rely on configuration files or programmatic definitions, making complex inter-component dependencies harder to reason about [6]. Some research has explored visual programming for distributed systems and dataflow models [12, 13], but these approaches often target specific domains or require significant user expertise in the underlying execution environment [8, 9]. Relatedly, FLaaS demonstrates practical cross-app federated learning on mobile devices, showing how distributed computation can be executed on real-world heterogeneous systems, though with a focus on model training rather than general workflow orchestration [14]. Studies on event-driven architectures highlight the benefits and challenges of reactive, asynchronous coordination patterns in distributed environments, including performance trade-offs [15]. However, these efforts rarely combine visual authoring with shared distributed memory and executable semantics in a unified model.

Other work has explored orchestration patterns and task scheduling for edge computing using DAG-based models, emphasizing the coordination of modular tasks across heterogeneous nodes [16]. Microservice-level orchestration frameworks illustrate multi-tier deployment strategies spanning edge and cloud environments, highlighting the challenges of distributed execution, resource management, and event coordination [17]. Approaches that extend dataflow paradigms across the edge-to-cloud continuum also explore similar structural principles to the graph model presented here, such as explicit task dependencies and runtime coordination [18]. Broader surveys of edge orchestration and distributed intelligence provide further context on the complexities of coordinating computation, communication, and state across heterogeneous infrastructures [19, 20].

The proposed editor distinguishes itself by combining a visual interface, explicit representation of shared distributed memory, and event triggers into a unified graph model [11]. This approach enables users to construct executable workflows without needing deep knowledge of the execution backend, while maintaining precise semantics for coordination and data sharing. Additionally, the ability to serialize graphs into machine-readable specifications bridges design-time modeling with runtime deployment, which is less common in existing solutions.

# 3 Graph Model and Semantics

This section describes the underlying graph model used by the proposed visual editor and defines the semantics of its core elements. The model is designed to capture computation, data sharing, and execution dependencies in a unified and executable representation suitable for edge and cloud computing environments.

## 3.1 Graph Overview

An application is modeled as a directed graph $G$=($V$,$E$) where each vertex $v \in V$ represents a functional entity and each directed edge $e \in E$ represents a dependency or interaction between entities. The graph is typed, meaning that vertices belong to distinct categories with well-defined semantics. The editor currently supports three primary vertex types: kernels, memory nodes, and event triggers.

The directed nature of the graph reflects both data flow and control flow, following principles of classical dataflow models [21, 22]. Depending on the connected vertex types, an edge may express data production/consumption, execution triggering, or coordination between kernels and shared memory. This explicit graph structure allows the application behavior to be reasoned about visually and interpreted programmatically.

In the proposed model, stored procedures, events, and event triggers are all instances of unikernels, differentiated by their execution semantics and triggering behavior, while data sources represent shared distributed data.

## 3.2 Kernel Nodes

Kernel nodes represent computational units within the application. A kernel corresponds to an externally defined executable entity, such as a function, service, or long-running process. The internal implementation of a kernel is outside the scope of the editor; instead, the editor focuses on how kernels are composed and coordinated.

Each kernel node is associated with a set of configurable properties, including but not limited to:

- A unique kernel identifier or name
- A reference to the external kernel implementation
- A list of arguments or configuration parameters

From a semantic perspective, a kernel node consumes data from incoming edges, performs computation, and may produce data or events on outgoing edges. Kernels may be stateless or stateful, depending on their interaction with memory nodes.

## 3.3 Memory Nodes

Memory nodes represent shared distributed memory used for data exchange and persistence between kernels. Unlike implicit data passing mechanisms, memory is modeled as a first-class graph element, making data sharing relationships explicit.

A memory node may be connected to multiple kernels, enabling many-to-many data access patterns. The semantics of a memory node include:

- A logical memory identifier
- Optional configuration parameters (e.g., access mode, consistency level)
- A mapping to an underlying distributed storage or memory backend at execution time

Edges connecting kernels to memory nodes define read and/or write access, depending on their direction and configuration. By explicitly representing memory in the graph, the editor enables clearer reasoning about shared state, data dependencies, and potential contention, supporting coordination patterns common in edge computing environments [15].

### 3.4 Event Trigger Nodes

Event trigger nodes model event-driven execution dependencies within the application. An event trigger represents a condition under which one or more kernels are activated. In the current design, events are primarily internal, such as the completion of a kernel or the availability of data in memory, although the model allows for future extension to user-defined or external events.

Event trigger nodes are associated with properties such as:

- Event type or identifier
- Trigger conditions
- Optional filtering or routing rules

Semantically, event triggers decouple the production of an event from the execution of kernels that respond to it. This supports flexible execution patterns common in edge and cloud environments, including asynchronous processing and reactive workflows [15].

### 3.5 Edges and Interaction Semantics

Edges in the proposed graph model are typed and encode not only connectivity but also data access semantics between nodes. In particular, the editor distinguishes between hard links and soft links, which define ownership and access rights over shared data.

A hard link indicates data ownership. When a kernel is connected to a memory node via a hard link, the kernel is considered the owner and producer of the data stored in that memory. The kernel is responsible for initializing, updating, and maintaining the consistency of the data. Other kernels may depend on this data, but ownership remains exclusively associated with the kernel connected through the hard link.

In contrast, a soft link represents read-only access. A kernel connected to a memory node via a soft link can consume or inspect the data but is not permitted to modify it. Soft links are used to model data dependencies where kernels rely on shared state without assuming responsibility for its lifecycle or consistency.

### 3.6 Executable Representation

The visual graph serves as an intermediate representation that can be automatically translated into a machine-readable specification, such as JSON or XML. This specification preserves both the structural topology of the graph and the semantic attributes of nodes and edges.

The resulting representation is passed to an execution API responsible for instantiating kernels, configuring shared distributed memory, and coordinating execution according to the defined dependencies. By maintaining a direct correspondence between visual elements and executable semantics, the proposed model bridges the gap between application design and deployment.

## 4 Visual Editor and User Interaction

This section describes the design and interaction model of the proposed visual editor. The editor is implemented as a web-based application and provides a domain-specific interface for authoring executable graphs that model kernel communication, shared memory usage, and event-driven execution.

### 4.1 Design Goals

The design of the editor is guided by the following goals:

- **Domain specificity**: provide abstractions tailored to kernel-based computation rather than generic diagramming.
- **Clarity of execution semantics**: make data sharing and execution dependencies explicit.
- **Low entry barrier**: enable users to construct valid graphs without requiring detailed knowledge of the underlying execution platform.
- **Direct executability**: ensure that every valid graph can be translated into an executable specification.

These goals influence both the visual notation and the interaction constraints enforced by the editor.

### 4.2 Visual Representation

The editor uses distinct visual shapes to differentiate node types:

- Stored procedures are represented as rectangles
- Events as rounded rectangles
- Event triggers as rounded rectangles highlighted in red to emphasize their control-flow role
- Data sources are rendered as cylinders, reflecting their function as shared distributed memory.

This visual differentiation allows users to quickly identify execution nodes and data-oriented components within the graph.

Directed edges are drawn as arrows connecting elements. Arrow direction encodes the direction of data flow or control dependency, depending on the connected node types. To visually distinguish data access semantics, hard links are rendered as solid lines, indicating data ownership, while soft links are rendered as dashed lines, representing read-only access to shared memory. The visual differentiation allows users to quickly interpret complex application graphs and identify execution paths. An example graph authored in the visual editor is shown in **Fig. 1**, illustrating the placement of kernels, memory nodes, and event triggers.

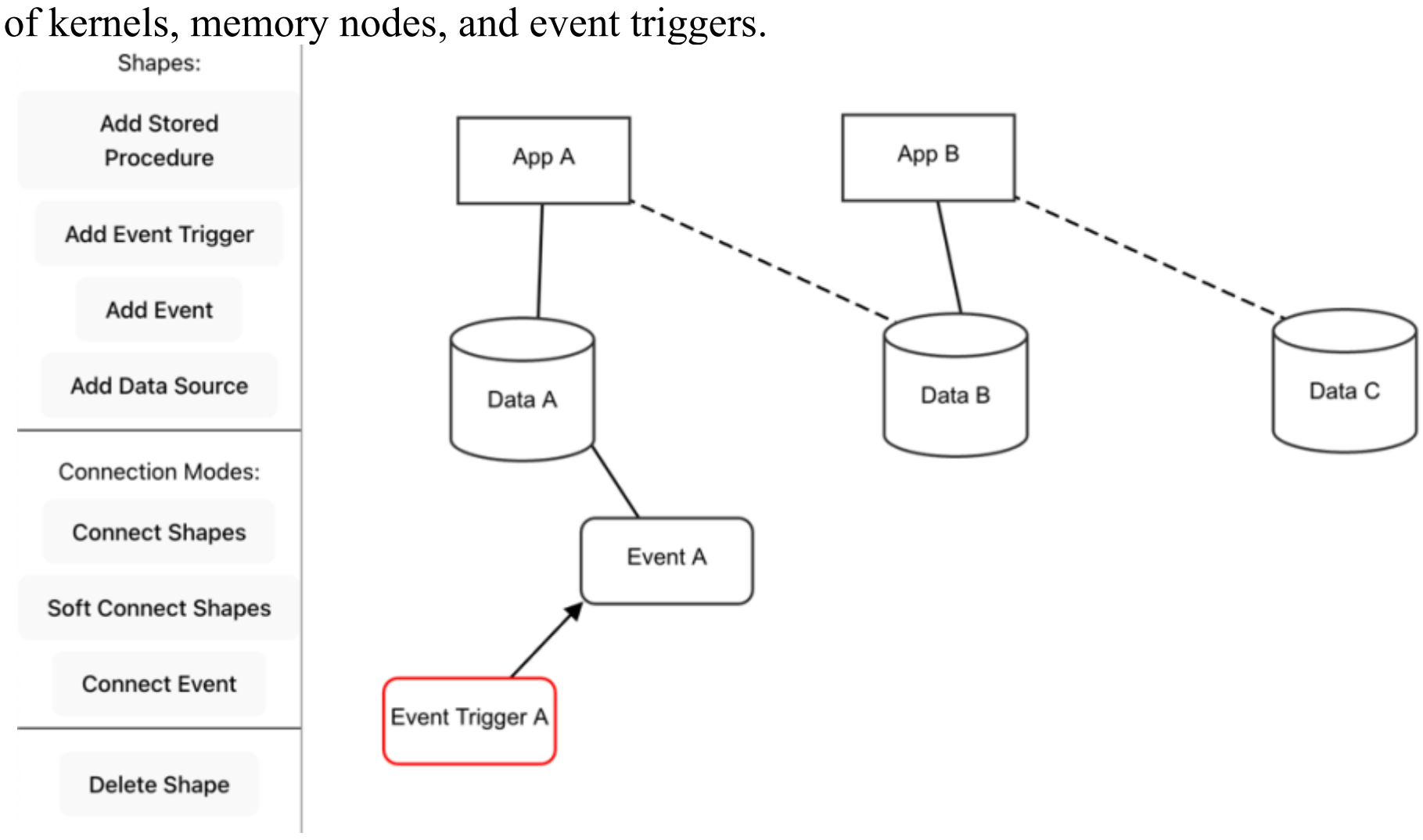

**Fig. 1.** Visual representation of a kernel communication graph. Rectangles denote stored procedures, rounded rectangles represent events, and red rounded rectangles indicate event triggers. Cylinders represent data sources. Solid and dashed edges denote hard and soft data links, respectively.

### 4.3 Interactive Graph Construction

Users construct graphs through direct manipulation. New elements can be added to the canvas and positioned freely, while edges are created by connecting compatible elements. The editor restricts invalid connections based on node types, preventing the creation of semantically incorrect graphs.

For example, kernels may be connected to memory nodes or event triggers, while event triggers serve as intermediaries for execution activation. Such constraints reduce modeling errors and ensure that the resulting graph remains executable.

### 4.4 Property Configuration

Each graph element exposes a property panel that allows users to define configuration parameters. Typical properties include:

- For kernels: name, reference to the external implementation, and argument definitions.

- For memory nodes: logical identifier and configuration options related to data access.
- For event triggers: event type and triggering conditions.

Properties are edited independently of the graph layout, enabling users to focus on structural design before refining configuration details. All properties are validated and stored as part of the graph model. Users can configure each node's properties through the panel, as illustrated in [**Fig. 2**].

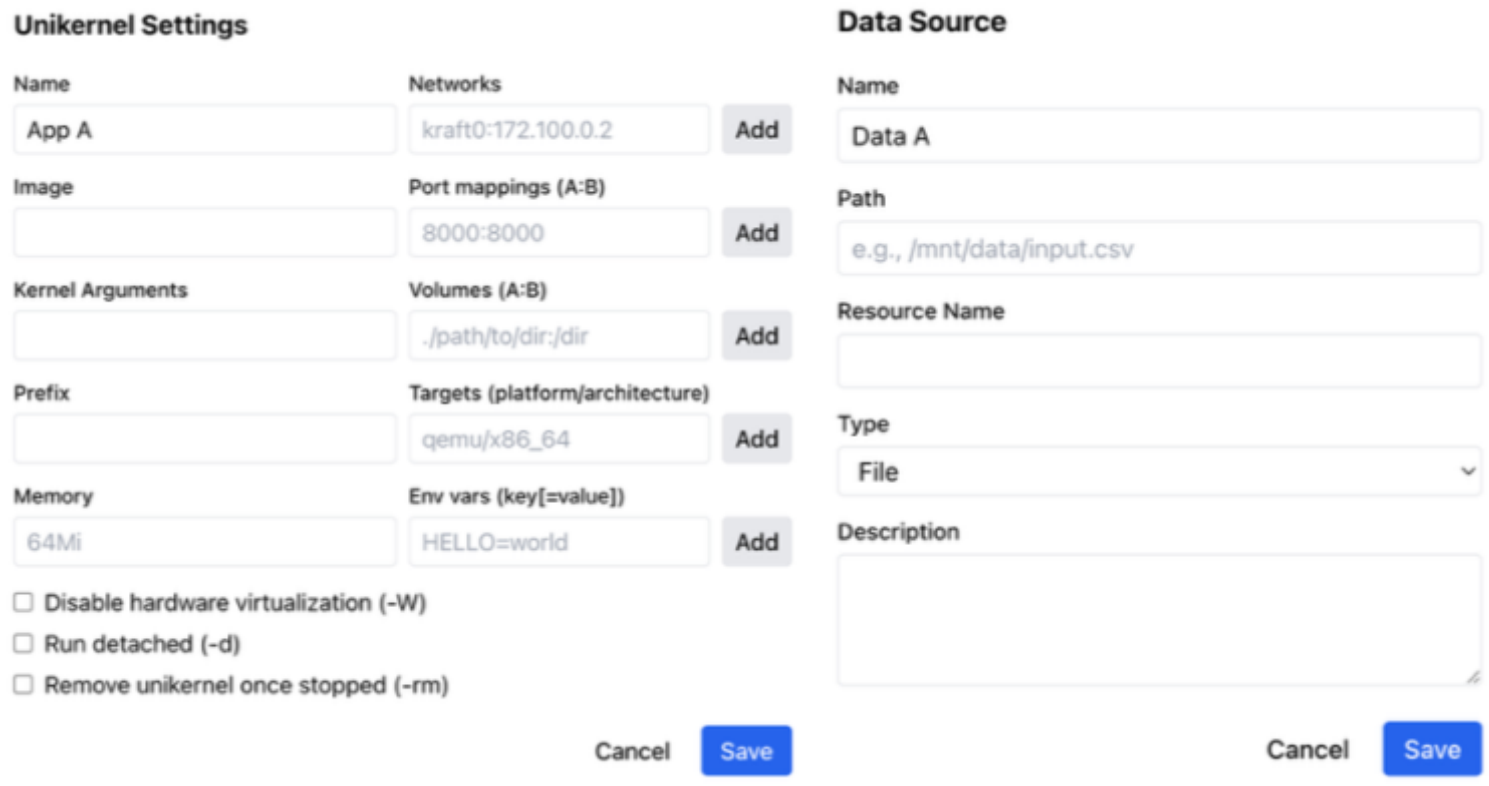

**Fig. 2**. Configuration dialog for a unikernel node (left), exposing execution parameters such as image selection, networking, memory limits, environment variables, and deployment targets. The same configuration model is reused for stored procedures, events, and event-trigger nodes and data source (right), allowing users to define resource identifiers, access paths, and metadata.

### 4.5 Validation and Consistency

The editor performs continuous validation to ensure graph consistency. This includes checking for missing required properties, invalid connections, and incomplete execution paths. Visual feedback is provided to guide users toward valid configurations.

By combining constrained interactions with real-time validation, the editor helps users' author complex kernel-based applications while preserving a clear mapping between the visual representation and its execution semantics.

## 5 Use Case: Federated Learning Orchestration

Federated learning is a distributed machine learning paradigm in which multiple clients collaboratively train a shared model without exchanging raw data. Instead, clients perform local training and periodically share model updates that are aggregated to produce a global model. This execution model is well suited to edge and cloud environments but introduces challenges related to coordination, data sharing, and event-driven execution [23, 24].

### 5.1 Motivation

Federated learning workflows are inherently modular and event-driven. They involve repeated interaction between local training processes, shared model state, and aggregation mechanisms. These interactions must be carefully orchestrated, particularly in heterogeneous environments where edge nodes may join or leave dynamically.

Traditional workflow tools often hide data sharing and execution dependencies, making it difficult to reason about the flow of model updates and control signals. This use case demonstrates how the proposed graph editor can be used to visually model a federated learning workflow with explicit semantics for computation, shared memory, and event triggers [24].

### 5.2 Graph-Based Federated Learning Model

In the proposed editor, a federated learning workflow is represented as a graph composed of kernel, memory, and event trigger nodes.

Kernel nodes represent computational tasks, such as:

- Local model training on edge devices
- Aggregation of model updates
- Evaluation of the global model

Memory nodes represent shared distributed state, including:

- The current global model parameters
- Aggregated model updates
- Optional metadata such as training rounds or convergence criteria

Event trigger nodes represent coordination events, such as:

- Completion of a local training round
- Availability of a sufficient number of client updates
- Initiation of a new federated learning round

Directed edges define how local training kernels read the global model from shared memory, write their updates, and trigger aggregation once updates are available. A federated learning workflow modeled in the editor is shown in [**Fig. 3**], demonstrating how kernels, shared memory, and event triggers interact during training.

### 5.3 Execution Flow

A typical execution cycle proceeds as follows. At the beginning of a training round, local training kernels are activated by an event trigger indicating the availability of the current global model. Each kernel reads the model parameters from shared distributed memory and performs training on local data.

Upon completion, local kernels write their model updates back to a shared memory node and emit an event signaling update availability. Once a predefined condition is met (e.g., a minimum number of updates), an aggregation kernel is triggered. This kernel reads the collected updates, computes a new global model, and writes the result back to shared memory.

The update of the global model triggers the next training round, closing the execution loop. This cyclic structure is naturally expressed in the graph representation and made explicit through visual connections between kernels, memory nodes, and event triggers.

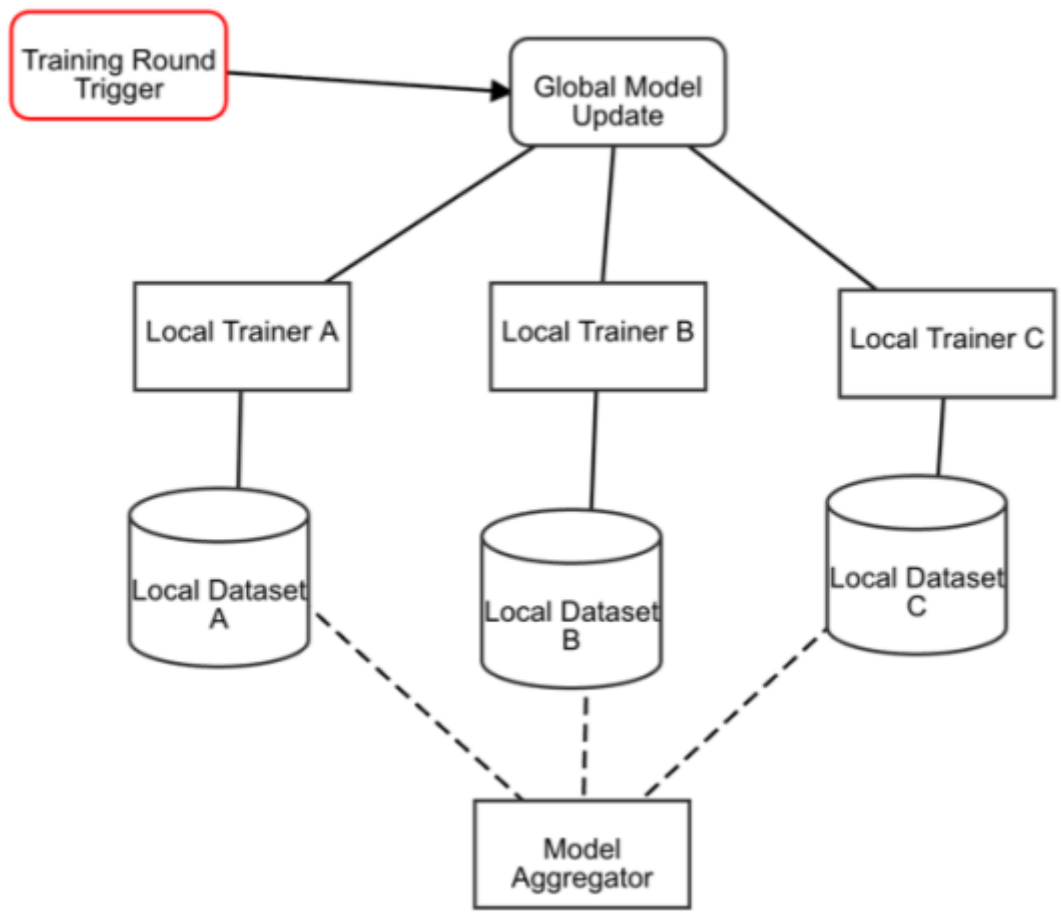

**Fig. 3.** Illustrates a federated learning workflow modeled using the proposed visual orchestration platform. Local training unikernels are hard-linked to private data sources, ensuring data locality, while soft links are used to share model updates with a central aggregation unikernel. This separation explicitly enforces privacy constraints at the orchestration level.

### 5.4 Benefits of the Visual Approach

Modeling federated learning workflows using the proposed editor provides several advantages:

- Explicit data sharing: shared model state is clearly represented, reducing ambiguity in how updates are exchanged.
- Transparent execution dependencies: event triggers make coordination logic visible and easier to reason about.
- Modularity: kernels can be independently modified or replaced without restructuring the entire workflow.
- Extensibility: additional components, such as validation or fault-handling kernels, can be added incrementally.

This use case illustrates how the editor supports complex, real-world distributed workflows and serves as a foundation for future experimental evaluation and deployment in edge and cloud environments, complementing prior federated learning frameworks [11, 23, 24, 25].

## 6 Conclusion and Future Work

This paper presented a domain-specific visual graph editor for modeling kernel-based applications in edge and cloud computing environments. The proposed system enables users to visually compose executable application graphs using first-class abstractions for computation, shared distributed memory, and event-driven execution. By explicitly modeling these elements and their interactions, the editor bridges the gap between application design and execution-oriented specification.

Unlike general-purpose diagramming tools, the proposed editor associates precise semantics with visual elements and enforces structural constraints that preserve executability. The resulting graphs can be automatically translated into machine-readable representations and passed to an execution API, providing a clear path from visual modeling to deployment. This approach supports modularity, improves reasoning about data sharing and execution dependencies, and lowers the barrier to designing complex distributed workflows.

While the current work focuses on the design of the graph model, user interface, and execution pipeline, several directions for future work remain. These include the implementation of a full execution backend, support for user-defined kernels and event types, and integration with existing edge and cloud orchestration frameworks. In addition, the system will be evaluated through real-world application scenarios, including federated learning workflows, where dynamic participation, shared model state, and event-driven coordination play a central role.

Overall, the proposed editor provides a foundation for visual orchestration of modular computation in distributed environments and opens new opportunities for exploring coordination, data sharing, and execution semantics in edge and cloud systems.

## Acknowledgment

This research has been supported by the Ministry of Science, Technological Development and Innovation (Contract No. 451-03-137/2025-03/200156) and the Faculty of Technical Sciences, University of Novi Sad through project "Scientific and Artistic Research Work of Researchers in Teaching and Associate Positions at the Faculty of Technical Sciences, University of Novi Sad 2025" (No. 01-50/295).